# Intrinsic breakdown strength: theoretical derivation and first-principles calculations

Shixu Liu,[†] Hongjun Xiang[†‡], Xin-Gao Gong[†‡], and Ji-Hui Yang[†‡]*

[†]Key Laboratory of Computational Physical Sciences (Ministry of Education), Institute of Computational Physics, Fudan University, Shanghai 200433, China

[‡]Shanghai Qi Zhi Institute, Shanghai 200230, China

*corresponding author

Email: jhyang04@fudan.edu.cn

## Abstract

Intrinsic breakdown strength ($F_{\text{bd}}$), as the theoretical upper limit of electric field strength that a material can sustain, plays important roles in determining dielectric and safety performance. The well accepted concept is that a larger band gap ($E_g$) often leads to a larger intrinsic breakdown strength. In this work, we analytically derive a simplified model of $F_{\text{bd}}$, showing a linear relationship between $F_{\text{bd}}$ and the maximum electron density of states ($\text{DOS}_{\max}$) within the energy range spanning from the conduction band minimum (CBM) to CBM + $E_g$. Using the Wannier interpolation technique to reduce the cost of calculating the $F_{\text{bd}}$ for various three- and two-dimensional materials, we find that the calculated $F_{\text{bd}}$ did not show any simple relationship with band gap, but it behaves linearly with the $\text{DOS}_{\max}$, consistent with our theoretical derivation. Our work shows that the $\text{DOS}_{\max}$ is more fundamental than the band gap value in determining the $F_{\text{bd}}$, thus providing useful physical insights into the intrinsic dielectric breakdown strength and opening directions for improving high-power devices. The dimensional effects on $F_{\text{bd}}$ has also been revealed that monolayers tend to have larger $F_{\text{bd}}$ due to reduced screening effects.

**KEYWORDS: intrinsic breakdown strength, first-principles, Wannier interpolation, electron-phonon coupling, maximum density of states, band gap, dimensional effects**

Breakdown strength is a critical parameter in the design of electrical devices and systems, playing a pivotal role in determining both the dielectric and safety performance [1–4]. As the dimensions of electrical devices continue to shrink, and the demand for higher power increases, there is a growing focus on materials with exceptional breakdown strength, capable of operating under extremely high electric fields. Materials such as SiC, GaN, and diamond have garnered substantial attention due to their potential applications in high-power devices [5–8]. Moreover, in the realm of energy storage, supercapacitors necessitate materials with high breakdown strength to achieve elevated energy densities [1,9–16]. Despite the notable advancements in this field, a comprehensive understanding of the fundamental mechanisms governing breakdown strength remains a crucial pursuit.

Under an external electric field, electrons in a dielectric material gain energy from the field, and then lose energy due to scatterings by phonons or collisions with ions. Depending on how electrons dissipate energy, there are various breakdown mechanisms such as intrinsic, electromechanical, thermal, and electrochemical breakdown. For an intrinsic breakdown process, electrons lose energy only through electron-phonon scattering. Therefore, the intrinsic breakdown strength (denoted as $F_{bd}$ hereafter) gives the upper boundary of the breakdown strength [17]. According to avalanche theory [18,19], at a low electric field, electrons have balanced rates of energy gain and loss. However, once the strength of the electric field exceeds the threshold, the average energy gain rate will become greater than the average energy loss rate, and the energy of electrons will continuously increase. As shown in Fig. 1, based on the von Hippel criterion [18], electrons that have an energy larger than the critical energy may collide with valence electrons, and excite them to the conductive bands, i.e., impact ionization. Consequently, the dielectric breakdown will happen. And for intrinsic breakdown, considering the energy conservation and momentum conservation, in direct band gap materials, the minimum energy required to excite a valence band electron to the conductive band is equal to the band gap ($E_g$) energy. For indirect band gap materials, this process may involve absorption or emission of phonons, but since phonon energy are generally much smaller than the band gaps of dielectric materials, they can be neglected while dealing with energy conservation. Therefore, the critical energy is considered to be $E_g$ here, which has been generally used in other works [20–22]. Note that, for some insulators that have band gaps larger than formation enthalpy ($\Delta H$), like LiF, the critical energy should be $\Delta H$ to account for the stability issue [22]. Intuitively, the larger the band gap, the larger the $F_{bd}$ [17,23,24]. Indeed, earlier models based on experimental data proposed a positive correlation between band gap and $F_{bd}$ [25,26], and recent theoretical studies also showed that band gap and phonon cutoff frequency are significantly related to the intrinsic breakdown field, although these studies used relatively coarse **q**-grids to sample the electron scattering [22,27]. The above insight has been a foundational consideration in the quest for increasing breakdown voltage and searching for dielectric materials with high breakdown strength [28–31]. Many works have tried to increase breakdown strength by increasing band gaps [12–14,32], however, the precise correlation between band gap values and intrinsic breakdown strength remains elusive. Consequently, a definitive consensus has yet to be reached on whether an increase in the band gap of a dielectric material invariably results in a commensurate enhancement of its intrinsic breakdown strength. Revealing this relationship is of great importance for gaining a comprehensive understanding of a material's performance under high electric field conditions, particularly for reliability in electronic devices and electrical equipment.

In this work, for the first time, we analytically derive that $F_{bd}$ is linearly dependent on the maximum electron density of states ($DOS_{max}$) within the energy range spanning from the CBM to the CBM + $E_g$ (for some materials is the CBM + $\Delta H$ due to the stability consideration) rather than the intuitive thought that $F_{bd}$ is positively correlated to band gap value. Using first-principles methods in combination with the Wannier interpolation technique to efficiently sample the electron-phonon scatterings on very dense **k**- and **q**-grids, we accurately calculate $F_{bd}$ for various systems ranging from three-dimensional crystals in different structures to two-dimensional (2D) monolayer

MoS₂ and phosphorene. For each system, we use different strains to tune its band gap as well as $DOS_{max}$. Our results show that, while the correlations between the $F_{bd}$ and band gap values behave differently for different systems and lack a universal trend, the dependence of the $F_{bd}$ on $DOS_{max}$ consistently behaves linearly for every system studied, without exception, thus demonstrating our model.

The energy gain rate of the electron with energy $E$ in a field $F$ can be written as [22]

$$A(E,F) = \frac{e^2 \tau(E) F^2}{3m^*} \quad (1),$$

where $e$ is the electronic charge and $m^*$ is the effective electronic mass. The average electron relaxation time $\tau(E)$ is the reciprocal of the average electron scattering rate, given by

$$\frac{1}{\tau(E)} = \frac{1}{D(E)} \sum_{n\mathbf{k}} \frac{1}{\tau_{n\mathbf{k}}} \delta(\varepsilon_{n\mathbf{k}} - E) \quad (2),$$

where $D(E)$ is the density of states (DOS), and $\tau_{n\mathbf{k}}$ describes the process in which an electron in the initial state $|n\mathbf{k}\rangle$ is scattered by a phonon with frequency $\omega_{\mathbf{q}\nu}$, reaching the final state $|m\mathbf{k}+\mathbf{q}\rangle$. Using Fermi's golden rule, $\tau_{n\mathbf{k}}$ can be evaluated as

$$\frac{1}{\tau_{n\mathbf{k}}} = \frac{2\pi}{\hbar} \sum_{m\mathbf{q}\nu} \left| g_{m\mathbf{k}+\mathbf{q},n\mathbf{k}}^{\mathbf{q}\nu} \right|^2 \left[ n_{\mathbf{q}\nu} \delta(\varepsilon_{n\mathbf{k}} - \varepsilon_{m\mathbf{k}+\mathbf{q}} + \hbar\omega_{\mathbf{q}\nu}) \right.$$
$$\left. + (n_{\mathbf{q}\nu} + 1) \delta(\varepsilon_{n\mathbf{k}} - \varepsilon_{m\mathbf{k}+\mathbf{q}} - \hbar\omega_{\mathbf{q}\nu}) \right] \quad (3),$$

where $n_{\mathbf{q}\nu}$ is the Bose-Einstein distribution for phonons and the delta function ensures energy conservation. The electron-phonon coupling (EPC) matrix element $g_{m\mathbf{k}+\mathbf{q},n\mathbf{k}}^{\mathbf{q}\nu}$ is given by

$$g_{m\mathbf{k}+\mathbf{q},n\mathbf{k}}^{\mathbf{q}\nu} = \sqrt{\frac{\hbar}{2M\omega_{\mathbf{q}\nu}}} |\langle \psi_{m\mathbf{k}+\mathbf{q}} | \xi_{\mathbf{q}\nu} \cdot \nabla_{\mathbf{R}} V_{\mathbf{q}} | \psi_{n\mathbf{k}} \rangle|^2 \quad (4),$$

where $M$ is the atomic mass, and $\psi$ is the wave function of the electron. The phonon polarization vector is $\xi_{\mathbf{q}\nu}$, and $\nabla_{\mathbf{R}} V_{\mathbf{q}}$ is the gradient of the potential with respect to collective atomic displacements from their equilibrium positions $\mathbf{R}$.

The energy loss rate of the electron during a phonon absorption or emission process can be evaluated by [22]

$$B(E) = \frac{2\pi}{\hbar D(E)} \sum_{n\mathbf{k}} \sum_{m\mathbf{q}\nu} (\hbar\omega_{\mathbf{q}\nu}) \left|g_{m\mathbf{k}+\mathbf{q},n\mathbf{k}}^{\mathbf{q}\nu}\right|^2 \left[(n_{\mathbf{q}\nu} + 1)\delta(\varepsilon_{n\mathbf{k}} - \varepsilon_{m\mathbf{k}+\mathbf{q}} - \hbar\omega_{\mathbf{q}\nu})\right.$$
$$\left. - n_{\mathbf{q}\nu}\delta(\varepsilon_{n\mathbf{k}} - \varepsilon_{m\mathbf{k}+\mathbf{q}} + \hbar\omega_{\mathbf{q}\nu})\right]\delta(\varepsilon_{n\mathbf{k}} - E) \quad (5).$$

Then, using the von Hippel breakdown criterion [18], i.e.,

$$A(E, F) > B(E), \quad E \in [\text{CBM}, \text{CBM} + E_g] \quad (6),$$

one can obtain

$$F_{\text{bd}} = \text{Max}\left[\frac{\sqrt{3m^*}}{e}\sqrt{\frac{1}{\tau(E)}B(E)}\right], \quad E \in [\text{CBM}, \text{CBM} + E_g] \quad (7).$$

As seen from Eqs. (1)-(7), the band gap value is not directly utilized in the calculation of $F_{\text{bd}}$. To determine the intrinsic breakdown field, we focus on the summation term of the electron energy by using an averaged phonon frequency and EPC strength to replace the momentum-dependent $\omega$ and $g$ terms. Note that the utilization of this approximation concept is prevalent in the field of superconductivity [33]. In this case, we rewrite Eqs. (2) and (5) (see the Supplemental Materials for the derivation [34]) as

$$\frac{1}{\tau(E)} = \frac{1}{\tau_{\mathbf{k}}} = \frac{2\pi}{\hbar} g_{\text{eff}}^2 (2n_0 + 1) D(E) \quad (8), \text{ and}$$

$$B(E) = \frac{2\pi}{\hbar}(\hbar\omega_{\text{mean}}) g_{\text{eff}}^2 D(E) \quad (9),$$

where $\hbar\omega_{\text{mean}}$ represents the average energy for phonons that participate in the scattering process, $n_0$ is the average occupation number of phonons, and $g_{\text{eff}}$ is the average EPC strength. As we can see, $\frac{1}{\tau(E)}$ and $B(E)$ are roughly proportional to $D(E)$. Accordingly, $F_{\rm bd}$ can be expressed as

$$F_{\rm bd} = \frac{\sqrt{3m^*}}{e}\frac{2\pi}{\hbar}g_{\rm eff}^2\sqrt{(2n_0+1)\hbar\omega_{\rm mean}}{\rm DOS}_{\rm max} \quad (10).$$

Thus, we have a linear relationship between $F_{\rm bd}$ and $\text{DOS}_{\rm max}$ if the remaining terms do not change much in a given material.

To demonstrate the above model, we calculate the intrinsic breakdown field strengths for various materials with diverse atomic structures, band structures and symmetries, including diamond and silicon in the diamond structure, ZnO (ZB-ZnO), AlAs, and SiC (ZB-SiC) in the zinc blende structure, GaN (W-GaN) in the wurtzite structure, LiF and NaCl in the NaCl-type structure, CsCl in the CsCl-type structure, as well as $MoS_2$ and phosphorene in the monolayer structure (m-$MoS_2$ and BP). The calculation details can be found in the Supplementary Materials [34]. To ensure calculation accuracy, relatively dense **k-** and **q**-grids are often needed for sampling in the Brillouin zone, which is very expensive. To reduce the computational cost, we first obtain necessary quantities on relatively sparse grids and then we adopt the Wannier interpolation technique to obtain band structures, phonon dispersions and EPC matrices on the fine grids. Convergence tests are conducted on these materials (see Supplementary Materials [34] for more details). In the following studies, we set the sizes of both **k-** and **q**-grids as $100 \times 100 \times 100$ for the bulk systems (for W-GaN, we use $100 \times 100 \times 60$ mesh), $300 \times 300 \times 1$ for m-$MoS_2$ and $320 \times 240 \times 1$ for BP. The smearing parameter is set at 0.01 eV. In general, the computational cost tends to increase rapidly with the density of the grid. Therefore, the utilization of the Wannier interpolation technique is indispensable and highly efficient for achieving a balance between accuracy and cost. Note that the electric field is assumed to be applied in the in-plane direction for 2D materials and BP exhibits strong anisotropic band structure near the CBM with the effective masses being 1.24 and 0.07 $m_{\rm e}$ along the zigzag and armchair directions, respectively. Therefore, we use the geometric mean value $m^* =$

$\sqrt{m_X^* m_Y^*}$ to estimate the effective mass in Eq. (7). The hydrostatic strain in bulk materials is applied by directly changing the lattice constant, which is defined as $(a - a_0)/a_0$, where $a$ and $a_0$ are the lattice constants with and without strain, respectively. While in 2D materials, biaxial strain is applied. To deal with the underestimation of band gaps under the Perdew-Burke-Ernzerhof exchange-correlation functional [36], we correct them using experimental values, i.e., $E_g^* = E_g - E_g^{\text{ref}} + E_g^{\text{exp}}$, where $E_g^{\text{exp}}$ is the experimental band gap, $E_g$ and $E_g^{\text{ref}}$ are the calculated band gaps with and without strain, respectively. Using our recently developed machine learning method [40], we also use the HSE functional [41] to get more accurate band structures and EPC matrices to calculate the $F_{\text{bd}}$.

Our calculated results are shown in Table I and Table SI. And the calculated $F_{\text{bd}}$ are in good agreement with previous theoretical results [22,27] and available experimental measurements. An interesting finding is that ZB-ZnO has an even smaller $F_{\text{bd}}$ than Si despite its much larger band gap, which completely violates the prevailing view. Note that, $F_{\text{bd}}$ is not only correlated to $E_g$ but also depends on the electron effective mass and EPC. To focus on revealing the correlation between $F_{\text{bd}}$ and $E_g$, we use strains to manipulate the bandgaps and study the dependence of $F_{\text{bd}}$ on bandgaps for each system. For bulk materials, the strains cover from -2% to +2%. For m-MoS$_2$, the strains cover from 0 to +2%, and for BP, the strains are in the range between -1% to 2%. Our calculation results are shown in Fig. 2. The results of more materials can be found in the Supplemental Materials [32]. For Si and BP, we find that the band gaps increase with increasing strain, which is consistent with previous results [51–55]. In contrast, for diamond [56], ZB-ZnO and m-MoS$_2$ [57], the bandgaps decrease with increasing strain. As for the relationship between bandgap and $F_{\text{bd}}$, we can see in Fig. 2(a)-(e) that only the results for Si and ZB-ZnO are consistent with the common view, whereas for other materials, the common view does not hold, especially for diamond and m-MoS$_2$.

In general, our results show that there is no clear dependence of $F_{bd}$ on band gap value, suggesting the absence of a universal trend.

Next, we examine the correlation between $F_{bd}$ and $DOS_{max}$. To eliminate the strain effects on effective masses and phonon frequencies, we normalize $DOS_{max}$ by multiplying by the $M$ factor, i.e., $M = \sqrt{m^*\omega_{mean}}$, for each material under different strains (see also Eq. 10) and we use $\omega_{cutoff}$ to approximate $\omega_{mean}$ here. The relationships between $MDOS_{max}$ and $F_{bd}$ are shown in Fig. 2(f)-(j). It is clearly seen that $F_{bd}$ is linearly dependent on the normalized $DOS_{max}$ for every system without exception, thus demonstrating our model. The slight deviation from linearity is mainly due to the variation of the $g_{eff}^2$ term under different strains.

To ascertain why the common view is flawed, we analyze the energy gain rates and energy loss rates for unstrained materials. As shown in Fig. 3, under a given electric field, electrons with low energies have higher energy gain rates compared to their energy loss rates. Consequently, the electron energy will increase until it reaches a pinning energy (PE) above which the energy loss rate exceeds the energy gain rate. For example, the PE for Si under an electric field of 63.9 MV/m is 0.95 eV, as shown in Fig. 3(a). Electrons with lower energies will undergo a net energy gain, while those with higher energies will experience a net energy loss. Thus, the average energy of electrons will be equal to the PE. With an increasing electric field, the PE also increases for Si, since the energy loss rate increases and the energy gain rate decreases monotonically with electron energy increasing from the CBM to the CBM + $E_g$, as seen in Fig. 3(a). Consequently, the PE also increases monotonically with the increasing electric field. When the PE reaches the maximum, that is, CBM + $E_g$, which is 1.17 eV for Si, conducting electrons will have sufficient energies to excite valance electrons into conduction bands, leading to breakdown. The increase in $E_g$ will lead to a corresponding increase in the PE maximum, thereby allowing for a larger electric field, that is, enhancement of $F_{bd}$. Therefore, the thought that a larger band gap leads to a larger $F_{bd}$ holds for Si. A similar situation exists for ZB-ZnO [see Fig. S2(a)], in which the PE maximum is also located at CBM + $E_g$ exactly under the breakdown field.

For diamond, however, we find that neither the energy loss rate nor the energy gain rate changes monotonically with electron energy increasing from the CBM to the CBM + $E_g$. As seen in Fig. 3(b), the energy loss rate has a peak and the energy gain rate has a dip around 4.3 eV. When the PE is smaller than 4.3 eV, the electron energies will be pinned by the PE and the breakdown will not happen. With further increase of the electric field, the PE of electrons increases. When the electric field reaches 2972.6 MV/m, the PE will reach 4.3 eV. Under this field, the energies of electrons will exceed 4.3 eV and keep increasing because the energy gain rate is always larger than the energy loss rate when the electron energy is larger than 4.3 eV. Ultimately, the electrons will have energies larger than the CBM + $E_g$, leading to breakdown. In this case, the $F_{bd}$ is not directly related to the $E_g$ but is determined by the PE maximum. As a larger $E_g$ does not necessarily lead to a larger PE maximum, the prevailing thought that a larger band gap leads to a larger $F_{bd}$ fails for diamond. Similar situations happen for m-MoS$_2$ and BP in which the PE maxima appear before CBM + $E_g$ [see Figs. S2(b) and S2(c)].

Note that, according to our model, the energy loss rate and gain rate reach the maximum and minimum, respectively, when the DOS reaches $DOS_{max}$ according to Eqs. (8) and (9). The consequence is that, with the increase of electric field strength before breakdown, the intersection will finally happen at the position of the $DOS_{max}$, which determines the PE maximum and thus the $F_{bd}$. For systems such as Si, the DOS increases monotonically with electron energy and reaches the maximum at the CBM + $E_g$. Therefore, if strain induces an increase in the band gap, it also results in a corresponding enhancement of $DOS_{max}$, leading to a larger $F_{bd}$. In this case, the prevailing view is valid. For systems such as diamond, the $DOS_{max}$ is reached before the electron energy reaches the CBM + $E_g$, as seen in Fig. S3(d). Although the tensile strain reduces the band gap, the $\text{DOS}_{\max}$ still increases. The consequence is that the $F_{\text{bd}}$ gets larger counterintuitively. In this case, the prevailing view fails. Our results show that, it is the $\text{DOS}_{\max}$ rather than the commonly believed band gap that plays the dominant role in determining $F_{\text{bd}}$.

Through our model, we can explain why ZB-ZnO has a smaller $F_{\text{bd}}$ than Si despite its larger experimental band gap of 3.27 eV [58] compared to 1.17 eV for Si. The band structure of ZB-ZnO shows that there is simply one band, mainly consisting of an s-orbital of Zn between the CBM and the CBM + $E_g$, as shown in Fig. S4. Hence, the $\text{DOS}_{\max}$ of ZB-ZnO is much smaller than that of Si, i.e., 0.22 v.s. 0.94, leading to the relatively low $F_{\text{bd}}$ in ZB-ZnO.

In addition, we can reveal the dimensional effects on the $F_{\text{bd}}$, since low-dimensional materials tend to suffer high electric field under operations. Generally speaking, dimensional effect is mainly manifested in the quantum confinement effect and reduction of screening effect. On the one hand, the quantum confinement effect will generally lead to the increase of the band gap values in monolayers and thus will affect the $\text{DOS}_{\max}$ between CBM and CBM + $E_g$. However, whether $\text{DOS}_{\max}$ will be larger or not in monolayers has no decisive conclusions. On the other hand, due to the reduced screening effect, EPC are often stronger in monolayers, which is good for obtaining larger $F_{\text{bd}}$. Taking $\text{MoS}_2$ as an example, we compare the breakdown field strength for monolayer and bulk phase in the Supplementary Materials [34]. We find that the calculated $F_{\text{bd}}$ of monolayer $\text{MoS}_2$ is larger than that of bulk phase, in agreement with experimental results [50]. However, the larger $F_{\text{bd}}$ in monolayer $\text{MoS}_2$ is actually mainly due to the stronger EPC rather than the increased band gap.

Finally, we want to mention that, to get more accurate $F_{\text{bd}}$, it is necessary to use accurate band structures and EPC matrices using more advanced functionals. We have discussed the calculated $F_{\text{bd}}$ using different functionals in the Supplementary Materials [34], and the results show that HSE functional can provide more reasonable results compared to experiments for some systems. Nevertheless, our model and our conclusion, which are not limited to the choice of special functionals, will not be quantitively affected.

In conclusion, we have developed an analytical model that shows a linear relationship between $F_{\text{bd}}$ and $\text{DOS}_{\text{max}}$ using appropriate approximations. To accurately determine the value of $F_{\text{bd}}$ within a reasonable time, we employed the Wannier interpolation technique in our calculations and showed the necessity of using dense **k**- and **q**-meshes to achieve convergence. By investigating the behaviors of $F_{\text{bd}}$ under different strains in different materials, we found that $F_{\text{bd}}$ has no clear dependence on band gap value and validated the effectiveness of our model even in cases where the prevailing view does not hold. Our work shows that the $\text{DOS}_{\text{max}}$ is more fundamental than the band gap value in determining the $F_{\text{bd}}$ and suggests that $F_{\text{bd}}$ might be enhanced by engineering the DOS using strategies like strain and doping, opening directions for improving high-power devices.

**ACKNOWLEDGMENTS**

This work was partially supported by the Guangdong Major Project of the Basic and Applied Basic Research (Future functional materials under extreme conditions 2021B0301030005), the China National Key R&D Program (2022YFA1404603), and the National Natural Science Foundation of China (Grants No. 12188101 and No. 11991061).

Table I. Calculated properties of Si, diamond, ZB-ZnO, m-MoS$_2$ and BP. The experimental values are obtained from ref [17,49,50,59–61].

| System | $a_0$ (Å) | $b_0$ (Å) | $E_g$ (eV) | $E_g^{exp}$ (eV) | $m^*$ ($m_e$) | $F_{bd}$ (MV/m) | $F_{bd}^{exp}$ (MV/m) |
|---|---|---|---|---|---|---|---|
| Si | 5.469 | | 0.608 | 1.17[a] | 0.959 | 79.9 | 50[e] |
| Diamond | 3.572 | | 4.173 | 5.48[a] | 1.642 | 2972.6 | 2150[e] |
| ZB-ZnO | 4.620 | | 0.699 | 3.27[b] | 0.128 | 71.0 | ---- |
| m-MoS$_2$ | 3.185 | | 1.658 | 1.90[c] | 0.447 | 225.8 | 433[f] |
| BP | 3.303 | 4.625 | 0.894 | 2.00[d] | 0.294 | 212.8 | ---- |

[a]Reference [59]
[b]Reference [60]
[c]Reference [49]
[d]Reference [61]
[e]Reference [17]
[f]Reference [50]

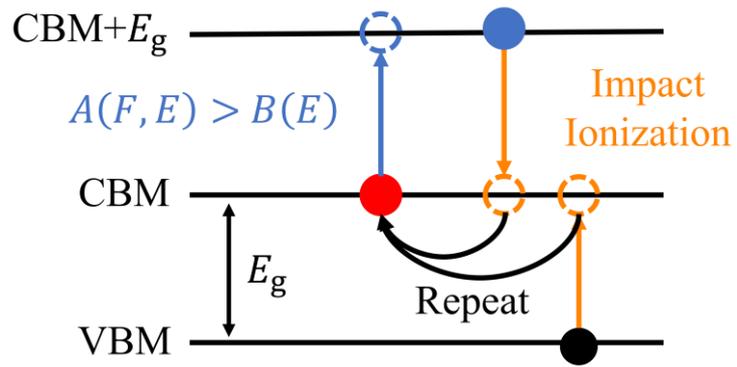

FIG. 1. Schematic diagram for the process of electron avalanche breakdown.

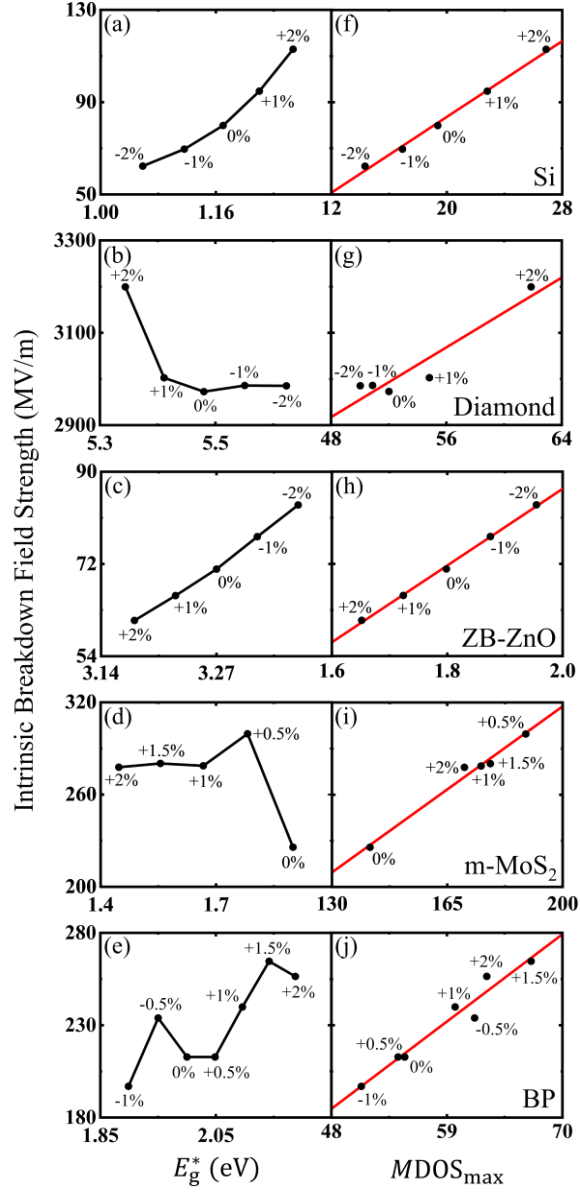

FIG. 2. Comparison of the relation between $F_{bd}$ and band gap with that between $F_{bd}$ and $MDOS_{max}$. Here, the band gap and $DOS_{max}$ are tuned using strains, which are labeled for each data. The calculated $F_{bd}$ as a function of $E_g^*$ for strained (a) Si, (b) diamond, (c) ZB-ZnO, (d) m-MoS$_2$ and (e) BP, respectively. The relationships between $F_{bd}$ and $MDOS_{max}$ for strained (f) Si, (g) diamond, (h) ZB-ZnO, (i) m-MoS$_2$ and (j) BP, respectively. The red lines are the results of linear fitting for eye guide. Note that, the calculated $F_{bd}$ did not follow a simple relationship with the band gap, while the linear dependence of $F_{bd}$ on the $MDOS_{max}$ holds for every system.

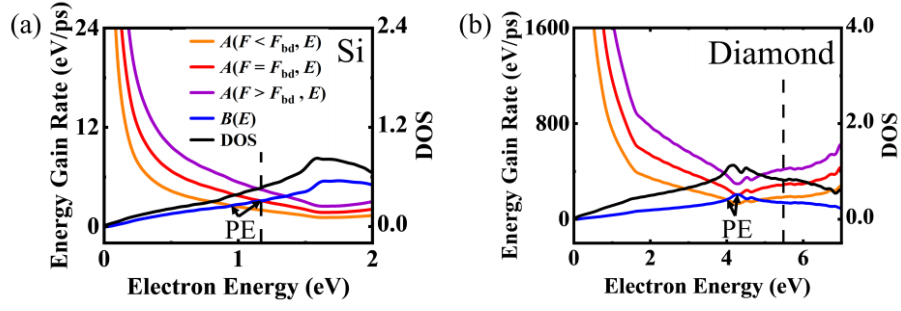

FIG. 3. The calculated breakdown properties for Si and diamond. The calculated energy gain rates, energy loss rates and DOS for (a) Si and (b) diamond. The red line is the energy gain rate under the external field of $F_{bd}$, while the yellow and purple line correspond to the energy gain rates when the external field is 20% smaller and larger than $F_{bd}$, respectively. The electron energy scale is referenced to the CBM and the vertical dotted lines represent the band gaps. Notably, the curve shapes of the energy gain rate and energy loss rate indicate that they are closely related to DOS, and the intersection of them does not necessarily locate at CBM + $E_g$ under the breakdown field.